\documentclass{PoS}

\usepackage{psfrag}

\newcommand{\be}{\begin{eqnarray}}
\newcommand{\ee}{\end{eqnarray}}
\newcommand{\ba}{\begin{array}}
\newcommand{\ea}{\end{array}}
\newcommand{\bi}{\begin{itemize}}
\newcommand{\ei}{\end{itemize}}

\def\beq{\begin{equation}}
\def\eeq{\end{equation}}
\newcommand{\eq}{\end{equation}}
\def\bea{\begin{eqnarray}}
\def\beqa{\begin{eqnarray}}
\def\eea{\end{eqnarray}}
\def\eqa{\end{eqnarray}}

\def\dv{\vec{\Delta}_t}
\def\ar{\alpha_\rho}

\title{Photoproduction of a large invariant mass $\gamma \rho$ pair at small momentum transfer}

\ShortTitle{Photoproduction of a large invariant mass $\gamma \rho$ pair at small momentum transfer}

\author{
\speaker{R.~Boussarie}\\
        LPT,   Universit\'{e} Paris-Sud, CNRS,  Universit\'e Paris-Saclay,  91405 Orsay, France \\
        E-mail: \email{Renaud.Boussarie@th.u-psud.fr}}
        
\author{B.~Pire \\
          Centre de Physique Th\'eorique, \'{E}cole Polytechnique, CNRS,  Universit\'e Paris-Saclay, 91128 Palaiseau, France \\
          E-mail: \email{bernard.pire@polytechnique.edu}
}

\author{L.~Szymanowski \\
          National Centre for Nuclear Research (NCBJ), Warsaw, Poland \\
          E-mail: \email{Lech.Szymanowski@ncbj.gov.pl}
}

\author{S.~Wallon \\
          LPT,   Universit\'{e} Paris-Sud, CNRS,  Universit\'e Paris-Saclay, 91405 Orsay, France \\
           UPMC, Universit\'{e} Paris 06, Facult\'{e} de Physique, 4 place Jussieu, 75252 Paris, France \\
          E-mail: \email{wallon@th.u-psud.fr}
}

\abstract{The collinear factorization framework allows to describe the exclusive photoproduction of a $\gamma\,\rho$ pair  in the generalized Bjorken regime in terms of a perturbatively calculable coefficient function and universal generalized parton distributions. The  kinematics are defined by   a large invariant mass of the $\gamma \rho$ pair and  a small transverse momentum of  the final nucleon. We  calculate the scattering amplitude at leading order in $\alpha_s$ and  the  differential cross sections for the process  where the $\rho-$meson is either longitudinally or transversely polarized, in the kinematics of the near future Jlab  experiments. 
 Our estimate of the cross section demonstrates that this
process is measurable at JLab~12-GeV.
}

\FullConference{XXIV International Workshop on Deep-Inelastic Scattering and Related Subjects, 11-15 April, 2016, DESY Hamburg, Germany}

\begin{document}

\section{Introduction and preliminaries}

We report here on a QCD calculation \cite{BoussarieProc} of the scattering amplitude for the process
\begin{equation}
\gamma(q) + N(p_1) \rightarrow \gamma(k) + \rho^0(p_\rho,\varepsilon(p_\rho)) + N'(p_2)\,.
\label{process1}
\end{equation}
Photoproduction of a pair of particles with large invariant mass 
is a natural case for using an extension of collinear QCD factorization theorems which describe the amplitudes
for deeply virtual
 Compton scattering (DVCS) and deeply virtual meson production~\cite{review}. The simplest example is timelike Compton scattering \cite{TCS} where the  produced particles are a lepton pair  with large invariant mass $Q$. In process (\ref{process1}), a wide angle Compton scattering subprocess $\gamma (q\bar q) \to \gamma \rho $ characterized by the large scale $M_{\gamma \rho}$ (the final state invariant mass) factorizes from generalized parton distribution (GPD). This  scale $M_{\gamma \rho}$ is related to the large transverse momenta transmitted to  the final photon and to  the final meson, but we insist that the final $\gamma \rho$ pair has an overall small transverse momentum (noted $\Delta_\perp$). Since the leading twist $\rho$ meson DAs have different chirality  when the $\rho$ is transversally or longitudinally polarized,  separating the  transverse  (resp. longitudinal) polarization of the $\rho$ meson allows one to get access to  chiral-odd (resp.  chiral-even) GPDs. This opens a new way to the 
extraction of GPDs and should thus constitute a useful check of their universality.

The study of such $2\to3$ processes was initiated in Ref.~\cite{IPST-eps}, where the process under study was the high energy photo (or electro) diffractive production of two vector mesons, the hard probe being the virtual "Pomeron" exchange (and the hard scale being the virtuality of this pomeron). A similar strategy has also been advocated in Ref.~\cite{Beiyad:2010cxa,kumano} to enlarge the number of processes which could be used to extract information on chiral-even GPDs.

The experimental study of  process (\ref{process1}) should not present major difficulties to modern detectors such as those developed for the 12 GeV upgrade of Jlab. The estimated rate depends of course much on the magnitude of GPDs.

\begin{figure}[h]

\psfrag{TH}{$T_H$}
\psfrag{Pi}{$\pi$}
\psfrag{P1}{$\,\phi$}
\psfrag{P2}{$\,\phi$}
\psfrag{Phi}{$\,\phi$}
\psfrag{Rho}{$\rho$}
\psfrag{tp}{$t'$}
\psfrag{s}{$s$}
\psfrag{x1}{\raisebox{-.1cm}{$\hspace{-.4cm}x+\xi$}}
\psfrag{x2}{\raisebox{-.1cm}{$\!x-\xi$}}
\psfrag{RhoT}{$\rho_T$}
\psfrag{t}{$t$}
\psfrag{N}{$N$}
\psfrag{Np}{$N'$}
\psfrag{M}{$M^2_{\gamma \rho}$}
\psfrag{GPD}{$\!GPD$}

\centerline{
\raisebox{1.6cm}{\includegraphics[width=13pc]{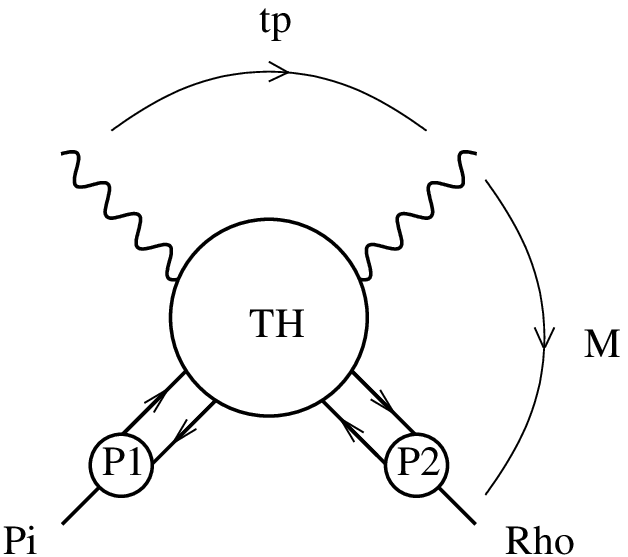}}~~~~~~~~~~~~~~
\psfrag{TH}{$\,  T_H$}
\includegraphics[width=13pc]{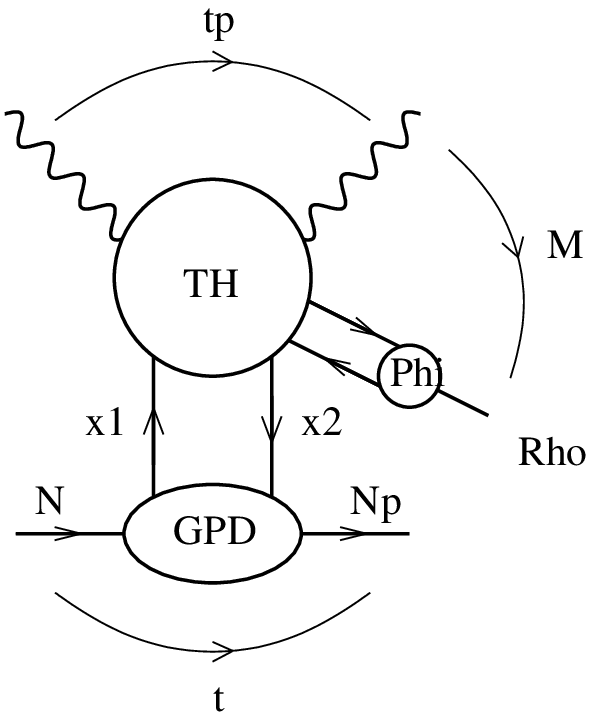}}

\caption{a) Factorization of the amplitude for the process $\gamma + \pi \rightarrow \gamma + \rho $ at large $s$ and fixed angle (i.e. fixed ratio $t'/s$); b) replacing one DA by a GPD leads to the factorization of the amplitude  for $\gamma + N \rightarrow \gamma + \rho +N'$ at large $M_{\gamma\rho}^2$\,.}
\label{Fig:feyndiag}
\end{figure}

The now classical proof of factorization of exclusive scattering at fixed angle and large energy~\cite{LB} allows to write the leading twist
 amplitude for the process $\gamma + \pi \rightarrow \gamma + \rho $ as the convolution of mesonic distribution amplitudes (DAs) of $\pi$ and $\rho$ and a hard scattering subprocess amplitude $\gamma  +( q + \bar q) \rightarrow \gamma + (q + \bar q) $ with the meson state replaced by a collinear quark-antiquark pair, as shown in Fig.~\ref{Fig:feyndiag}a. 
We extend the strategy used in the factorization procedure of the exclusive meson electroproduction amplitude near the forward region~\cite{fact} by replacing in Fig.~\ref{Fig:feyndiag}a the lower left meson DA by a $N \to N'$ GPD, and thus get Fig.~\ref{Fig:feyndiag}b. Indeed the same collinear factorization property bases the validity of the leading twist approximation, which either replaces the meson wave function by its DA or describes the $N \to N'$ transition through GPDs. A slight difference is that light-cone fractions ($z, 1- z$) leaving the DA are positive, but the corresponding fractions ($x+\xi,\xi-x$) may be positive or negative in the case of the GPD. Our calculation  shows that this difference does not spoil the factorization property, at least at the (leading) order that we are working here.

Let us briefly specify our set of conventions. We define
$
P^\mu = \frac{p_1^\mu + p_2^\mu}{2} ~,~ \Delta^\mu = p_2^\mu - p_1^\mu\,,
$
and decompose momenta on a Sudakov basis  (with $p$ and $n$ the light-cone vectors and $2p\cdot n = s $) as
\beqa
 p_1^\mu &=& (1+\xi)\,p^\mu + \frac{M^2}{s(1+\xi)}\,n^\mu~, \quad p_2^\mu = (1-\xi)\,p^\mu + \frac{M^2+\vec{\Delta}^2_t}{s(1-\xi)}n^\mu + \Delta^\mu_\bot\,, \quad q^\mu = n^\mu ~,\\
k^\mu &=& \alpha \, n^\mu + \frac{(\vec{p}_t-\vec\Delta_t/2)^2}{\alpha s}\,p^\mu + p_\bot^\mu -\frac{\Delta^\mu_\bot}{2}~,
 p_\rho^\mu = \alpha_\rho \, n^\mu + \frac{(\vec{p}_t+\vec\Delta_t/2)^2+m^2_\rho}{\alpha_\rho s}\,p^\mu - p_\bot^\mu-\frac{\Delta^\mu_\bot}{2}\,,\nonumber 
\eqa
with 
$M$, $m_\rho$ the nucleon and  $\rho$ meson masses.
The total cms energy squared of the $\gamma$-N system is
$
S_{\gamma N} = (q + p_1)^2 = (1+\xi)s + M^2\,.
$
On the nucleon side, the squared transferred momentum is
\begin{equation}
\label{transfmom}
t = (p_2 - p_1)^2 = -\frac{1+\xi}{1-\xi}\vec{\Delta}_t^2 -\frac{4\xi^2M^2}{1-\xi^2}\,.
\end{equation}
The other useful Mandelstam invariants read
\begin{eqnarray}
\label{M_pi_rho}
s'&=& ~(k +p_\rho)^2 = ~M_{\gamma\rho}^2= 2 \xi \, s \left(1 - \frac{ 2 \, \xi \, M^2}{s (1-\xi^2)}  \right) - \dv^2 \frac{1+\xi}{1-\xi}\,, \\
\label{t'}
- t'&=& -(k -q)^2 =~\frac{(\vec p_t-\vec\Delta_t/2)^2}{\alpha} \;,\\
\label{u'}
- u'&=&- (p_\rho-q)^2= ~\frac{(\vec p_t+\vec\Delta_t/2)^2+(1-\alpha_\rho)\, m_\rho^2}{\alpha_\rho}
 \; .
\end{eqnarray}

The hard scale $M^2_{\gamma\rho}$ is the invariant squared mass of the ($\gamma$ $\rho^0$) system.  The leading twist calculation of the hard part only involves the approximated kinematics in the generalized Bjorken limit: neglecting $\vec\Delta_\bot$ in front of $\vec p_\bot$ as well as hadronic masses, it amounts to
\beqa
\label{skewness2}
M^2_{\gamma\rho} \approx  \frac{\vec{p}_t^2}{\alpha\bar{\alpha}} 
~;~
\ar \approx 1-\alpha =\bar \alpha~;~
\xi =2  \frac{\tau}{2-\tau} ~,\tau \approx 
\frac{M^2_{\gamma\rho}}{S_{\gamma N}-M^2} ~;
-t'  \approx  \bar\alpha\, M_{\gamma\rho}^2  ~,-u'  \approx  \alpha\, M_{\gamma\rho}^2 \,.
\eqa

\section{The hard amplitude}

The computation of the scattering amplitude of the process (\ref{process1}) follows the usual procedure; it reads
\beq
\label{AmplitudeFactorized}
\mathcal{A}(t,M^2_{\gamma\rho},p_T)  =\frac{1}{\sqrt{2}} \int_{-1}^1dx\int_0^1dz\ (T^u(x,z) \, F^{u}(x,\xi,t)    -T^d(x,z)\, F^{d}(x,\xi,t)) \,\Phi_{||,\bot}(z)\,,
\eq
\def\diagici{2.65cm}
\begin{figure}[h!]
\begin{center}
\psfrag{z}{\begin{small} $z$ \end{small}}
\psfrag{zb}{\raisebox{0cm}{ \begin{small}$\bar{z}$\end{small}} }
\psfrag{gamma}{\raisebox{+.1cm}{ $\,\gamma$} }
\psfrag{pi}{$\,\pi$}
\psfrag{rho}{$\,\rho$}
\psfrag{TH}{\hspace{-0.2cm} $T_H$}
\psfrag{tp}{\raisebox{.5cm}{\begin{small}     $t'$       \end{small}}}
\psfrag{s}{\hspace{.6cm}\begin{small}$s$ \end{small}}
\psfrag{Phi}{ \hspace{-0.3cm} $\phi$}
\hspace{-.2cm}
\begin{tabular}{ccccc}
\includegraphics[width=\diagici]{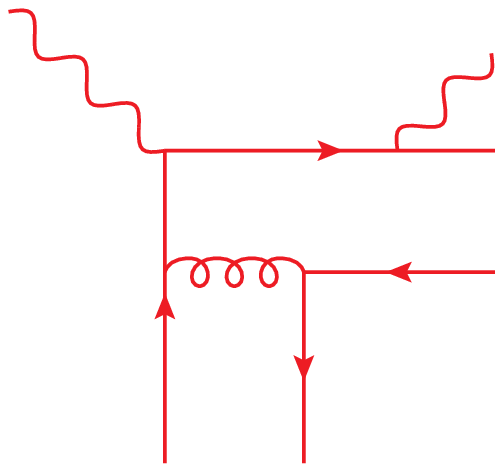}
&
\includegraphics[width=\diagici]{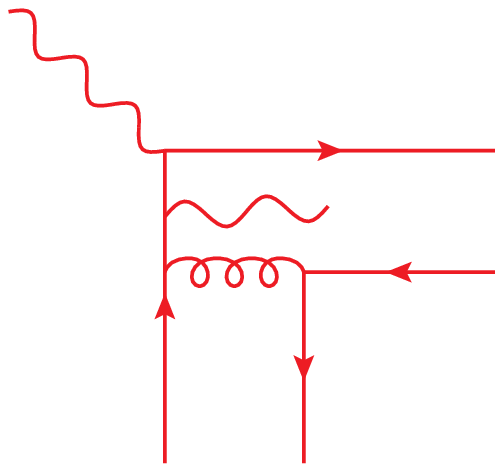}
&
\includegraphics[width=\diagici]{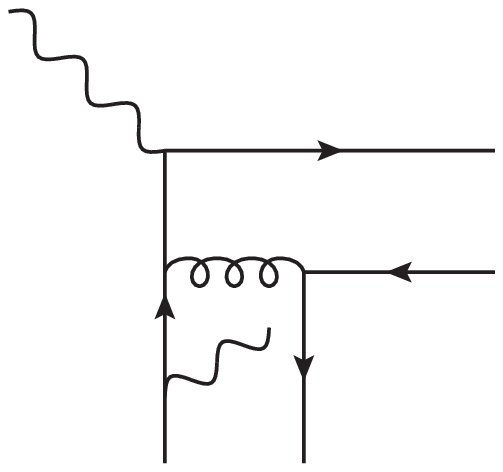}
&
\includegraphics[width=\diagici]{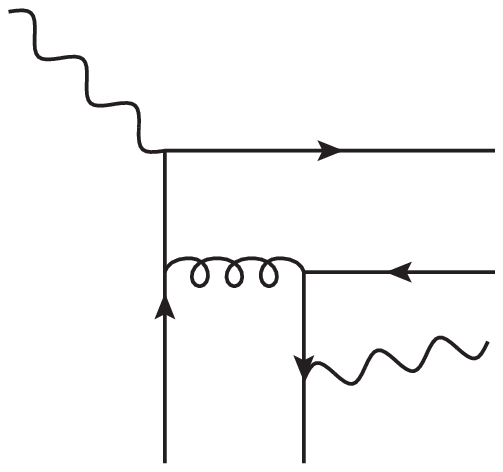}
&
\includegraphics[width=\diagici]{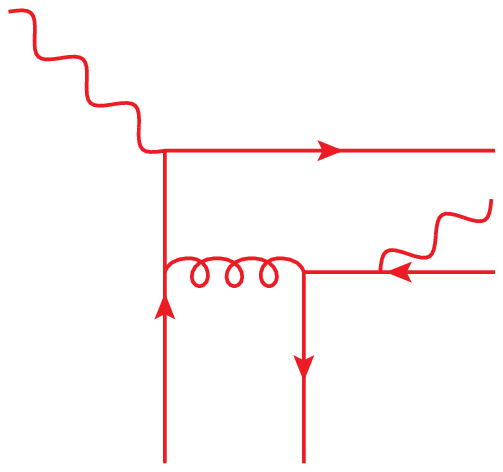} \\
$A_1$ & $A_2$ & $A_3$ & $A_4$ & $A_5$ \\
\\
\includegraphics[width=\diagici]{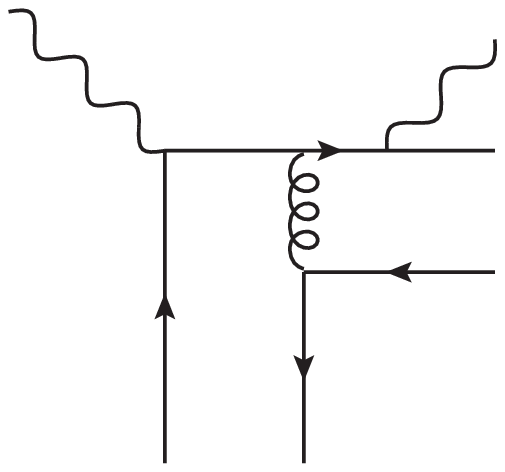}
&
\includegraphics[width=\diagici]{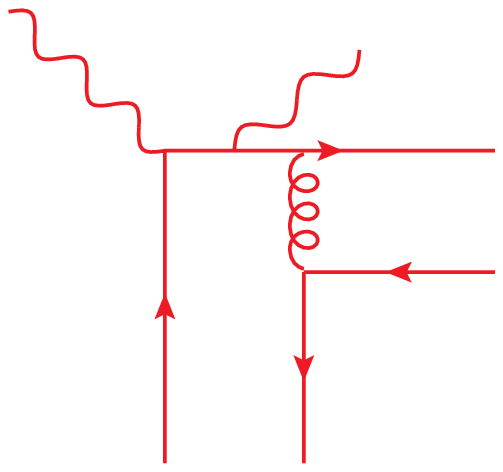}
&
\includegraphics[width=\diagici]{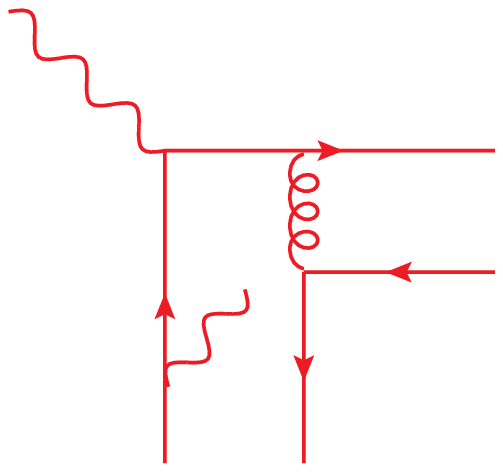}
&
\includegraphics[width=\diagici]{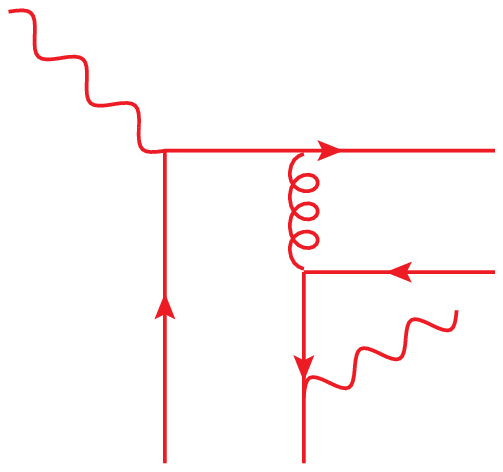}
&
\includegraphics[width=\diagici]{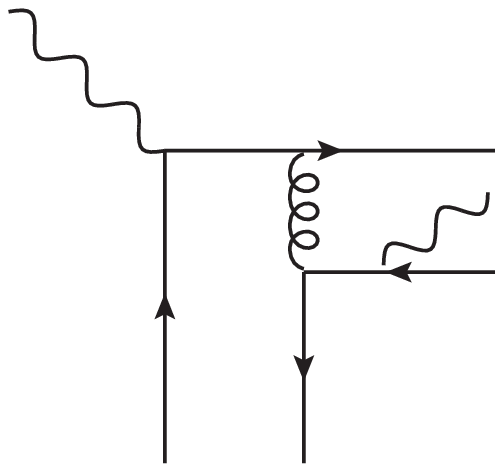}
\\
$B_1$ & $B_2$ & $B_3$ & $B_4$ & $B_5$
\end{tabular}
\caption{Half of the Feynman diagrams contributing to the hard amplitude. In the chiral-odd case, $A_3$, $A_4$ and $B_1$, $B_5$ are the only contributing diagrams (the red diagrams cancel in this case).}
\label{Fig:diagrams}
\end{center}
\end{figure}
%
%
where $T^u$ and $T^d$ are the hard parts of the amplitude where the photon couples respectively to a $u$-quark  and to a $d$-quark and $F^{u,d}$ is a chiral-odd or chiral-even GPD. This decomposition, with the $\frac{1}{\sqrt{2}}$ prefactor, takes already into account that the  $\rho^0$-meson is described as $\frac{u\bar{u}-d\bar{d}}{\sqrt{2}}$.
 In total there are 20 diagrams which can contribute to both processes at twist-2 level. However in the case of  the $(\gamma, \rho_\bot )$ production, due to the chiral-odd structure of the DA and of the GPD, only 8 diagrams do not vanish. But it is enough to calculate 4 out of
those 8 diagrams, for example the four black diagrams of Fig.~\ref{Fig:diagrams}, and to obtain the contribution of the remaining 4 diagrams using symmetry properties.
In the case of   $(\gamma, \rho_L )$ production, all 20 diagrams contribute, but similarly as above, it is enough to calculate only 10 of them, shown in Fig.~\ref{Fig:diagrams}, 
 since the contribution of the remaining 10 diagrams can be obtained using symmetry properties. 

\section{Results}

%
%
\psfrag{T}{} 
\def\sca{.7}
\psfrag{0}{\scalebox{\sca}{$0$}}
\psfrag{1}{\scalebox{\sca}{$1$}}
\psfrag{2}{\scalebox{\sca}{$2$}}
\psfrag{3}{\scalebox{\sca}{$3$}}
\psfrag{4}{\scalebox{\sca}{$4$}}
\psfrag{5}{\scalebox{\sca}{$5$}}
\psfrag{6}{\scalebox{\sca}{$6$}}
\psfrag{7}{\scalebox{\sca}{$7$}}
\psfrag{8}{\scalebox{\sca}{$8$}}
\psfrag{9}{\scalebox{\sca}{$9$}}
\psfrag{.}{\scalebox{\sca}{$.$}}
\begin{figure}[!h]
\begin{center}
\psfrag{H}{\hspace{-1.5cm}\raisebox{-.6cm}{\scalebox{.7}{$-u' ({\rm GeV}^{2})$}}}
\psfrag{V}{\raisebox{.3cm}{\scalebox{.7}{$\hspace{-.4cm}\displaystyle\frac{d \sigma_{\rm even}}{d M^2_{\gamma \rho} d(-u') d(-t)}~({\rm nb} \cdot {\rm GeV}^{-6})$}}}
\includegraphics[width=7cm]{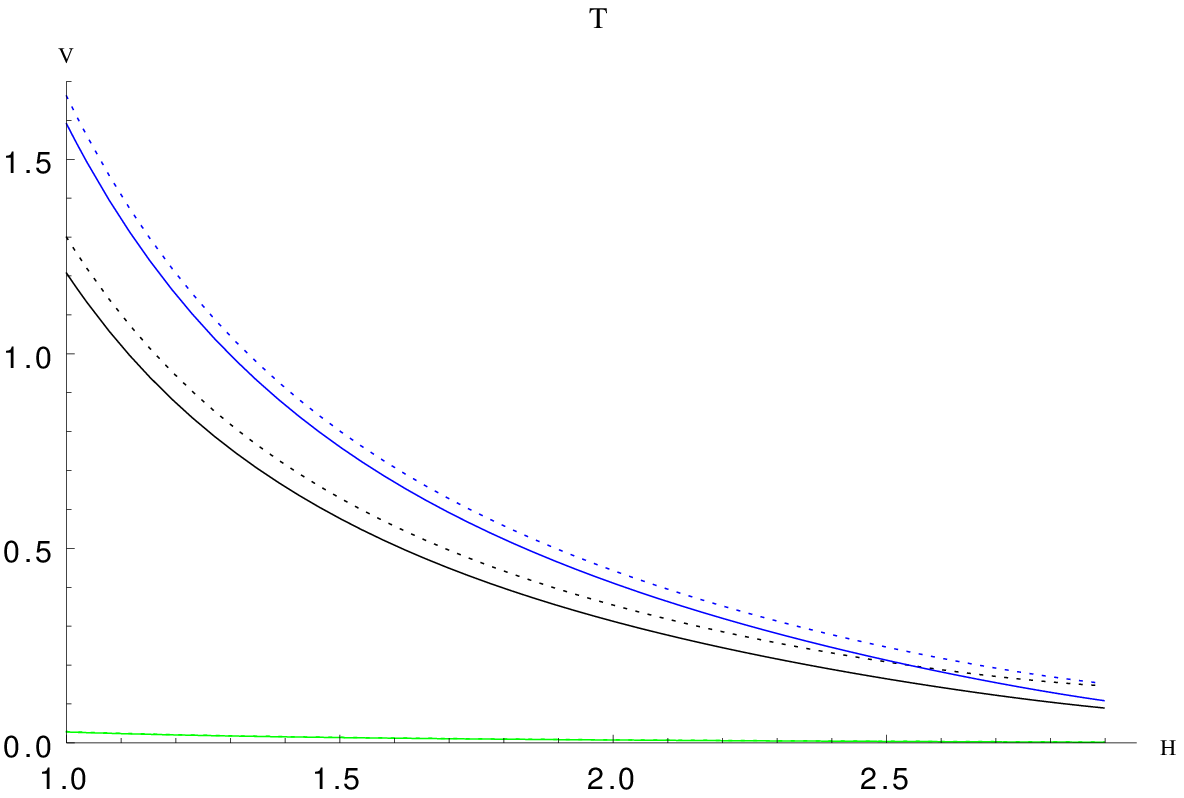}
\psfrag{H}{\hspace{-1.5cm}\raisebox{-.6cm}{\scalebox{.7}{$M^2_{\gamma \rho}~({\rm GeV}^{2})$}}}
\psfrag{V}{\raisebox{.3cm}{\scalebox{.7}{$\hspace{-.4cm}\displaystyle\frac{d\sigma_{even}}{d M^2_{\gamma\rho}}~({\rm nb} \cdot {\rm GeV}^{-2})$}}}
\raisebox{-.13cm}{\includegraphics[width=7.2cm]{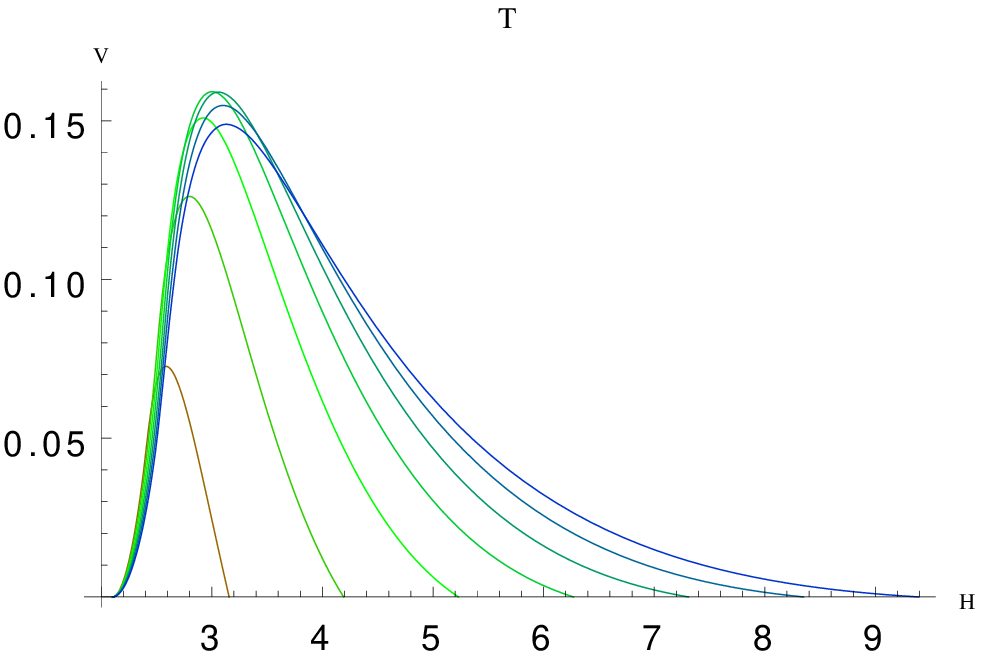}}
\caption{the differential cross section for the photo production of a photon and a longitudinal $\rho$ on a proton target. 
Left: $u'$ dependence of $\frac{d\sigma}{dt \,du' \, dM^2_{\gamma\rho}}$ for $s_{\gamma N} = 20~GeV^2$, $-t=(-t)_{min}$ and $M^2_{\gamma\rho}= 4~GeV^2$ (black lines) and shows the dominance of the $u-$quark contribution (blue lines) over the $d-$quark contribution (green lines) due to the charge effect. Note that the interference between $u-$quark and $d-$quark contributions is important and negative. Solid and dotted lines refer to two slightly different models of sea quark polarized PDFs. Right: $M^2_{\gamma\rho}$ dependence of $\frac{d\sigma}{dM^2_{\gamma\rho}}$ for values of $S_{\gamma N}$ in the set 8, 10, 12, 14, 16, 18, 20 ${\rm GeV}^{2}$ (from 8: left, brown to 20: right, blue), covering the JLab~12-GeV energy range.}
\label{Fig:dsigmaEVEN}
\end{center}
\end{figure}

The differential cross-section 
$\left.\frac{d\sigma}{dt \,du' \, dM^2_{\gamma\rho}}\right|_{\ -t=(-t)_{min}} = \frac{|\mathcal{M}|^2}{32S_{\gamma N}^2M^2_{\gamma\rho}(2\pi)^3},$ 
a function of ($M^2_{\gamma\rho},u'$), 
is dominated by its chiral-even part.
As an example of our results, we show on the left panel of Fig.~\ref{Fig:dsigmaEVEN}  the cross-section in the specific kinematics  $M^2_{\gamma \rho} = 4$~GeV$^2$, $S_{\gamma N}= 20$~GeV$^2$ as a function of $-u'$. We show the relative contributions of the $u-$ and $d-$quark GPDs (adding the vector and axial contributions), which interfere in a destructive way because of the flavor structure of the $\rho^0 = \frac{u \bar u - d \bar d}{\sqrt 2}$. Introducing a phenomenological $t$-dependence and integrating over $t$ and $u'$ in a sensible range, we get the less differential cross section $\frac{d\sigma}{dM^2_{\gamma\rho}}$  shown on the  right panel of Fig.~\ref{Fig:dsigmaEVEN}   for various values of $S_{\gamma N}$ in the range accessible at JLab~12-GeV.
The order of magnitude of the cross-section demonstrate that this experiment should be feasible at JLab~12-GeV, thus opening a new playground for the extraction of GPDs and a most useful test of their universality. 

\section*{Acknowledgments}

This work is partly supported by grant No 2015/17/B/ST2/01838 by the National Science Center in Poland, by the French grant ANR PARTONS (Grant No. ANR-12-MONU-0008-01), by the COPIN-IN2P3 agreement, by the Labex P2IO and by the Polish-French collaboration agreement Polonium.

\end{document}